\documentstyle[epsfig,color,12pt]{article}

\include{epsf}

\def\beq{\begin{equation}}
\def\eeq{\end{equation}}

\begin{document}

\centerline{\large\bf Electromagnetic source for the Kerr-Newman
geometry}

\vskip 0.2in

\centerline{\bf Irina~Dymnikova}

\vskip 0.2in

\centerline{\it A.F. Ioffe Physico-Technical Institute,
Politekhnicheskaja 26, St.Petersburg, 194021 Russia}

\centerline{\it Department of Mathematics and Computer Science,
University of Warmia and Mazury,}

\centerline{\it S{\l}oneczna 54, 10-710  Olsztyn, Poland; e-mail:
irina@uwm.edu.pl}

\vskip0.2in

{\bf Abstract}

\vskip0.1in

Source-free equations of nonlinear electrodynamics minimally
coupled to gravity admit regular axially symmetric asymptotically
Kerr-Newman solutions which describe electrically charged rotating
black holes and spinning solitons. Asymptotic analysis of
solutions shows the existence of de Sitter vacuum interior which
has the properties of a perfect conductor and ideal diamagnetic.
The Kerr ring singularity (a naked singularity in the case without
horizons) is replaced with a superconducting current which serves
as a non-dissipative source of the Kerr-Newman fields and can be
responsible for an unlimited life time of a spinning object.


\vskip0.1in

{\bf Journal Reference: Intern. J. Mod. Phys. D Vol. 24, No. 14
(2015) 1550094}

\vskip0.2in

The Kerr-Newman solution to the Maxwell-Einstein equations was
obtained in 1965 \cite{newman}
  $$
ds^2 = \biggl(\frac{2mr-e^2}{\Sigma} - 1\biggr) dt^2  +
\frac{\Sigma}{\Delta} dr^2  -
\frac{2a(2mr-e^2)\sin^2\theta}{\Sigma}dt d\phi
 $$
 $$
+ \Sigma d\theta^2 + \biggl(r^2 + a^2
+ \frac{(2mr-e^2)a^2\sin^2\theta}{\Sigma}\biggr)\sin^2\theta
d\phi^2
                                                                        \eqno(1)
  $$
with using the Newman-Janis algorithm \cite{trick}.
Here
  $$
 \Sigma = r^2 + a^2\cos^2\theta;
  ~~ \Delta = r^2 - 2mr + a^2 + e^2  .
                                                         \eqno(2)
 $$
 The associated electromagnetic
potential is given by \cite{newman} $A_{\mu} = - (e r/\Sigma)[1;
0, 0, -a\sin^2\theta]$.

In 1968 Carter found that the
parameter $a$ couples with the mass $m$ to give the angular
momentum $J=ma$ and independently couples with the charge $e$ to
give an asymptotic magnetic dipole moment $\mu=ea$, so that the
gyromagnetic ratio $e/m$ is exactly the same as predicted for a
spinning particle by the Dirac equation \cite{carter}.

Carter discovered also that in the case appropriate for a
particle, $a^2+e^2>m^2$, when there are no Killing horizons and
the manifold is geodesically complete (except for geodesics which
reach the singularity),  any point can be connected to any other
point by both a future and a past directed time-like curve which
originate in the region where $g_{\phi\phi} <0$, can extend over
the whole manifold and cannot be removed by taking a covering
space \cite{carter}.

The source models for the Kerr-Newman fields, involving a
screening or covering of causally dangerous region, can be divided
into disk-like \cite{werner,bur74,hamity,lopez1}, shell-like
\cite{delacruz,cohen,boyer1,lopez}, bag-like
\cite{boyer,trumper,tiomno,bur89,behm,bur2015}, and string-like
(\cite{bur2010,bur2013} and references therein). The problem of
matching the Kerr-Newman exterior to a rotating material source
does not have a unique solution, since one is free
 to choose arbitrarily the boundary between the exterior
and the interior \cite{werner} as well as an interior model.


The problem of a regular source for the Kerr-Newman
electromagnetic fields can be approached in the frame of nonlinear
electrodynamics coupled to gravity (NED-GR)\footnote{NED theories
appear as low-energy effective limits in certain models of
string/M-theories \cite{fradkin,tseytlin,witten}.}.

Nonlinear electrodynamics was proposed by Born and Infeld in 1934
as founded on two basic points: to consider electromagnetic field
and particles within the frame of one physical entity which is
electromagnetic field; to avoid letting physical quantities become
infinite \cite{born}. In their theory particles are considered as
singularities of the field, with mass of electromagnetic origin,
but it is also possible to obtain the finite electron radius by
introducing an upper limit on the electric field. In both cases a
total energy is finite \cite{born}.

The Born-Infeld program can be realized in nonlinear
electrodynamics minimally coupled to gravity. Source-free NED-GR
equations admit regular causally safe axially symmetric
asymptotically Kerr-Newman solutions \cite{me2006}, which describe
regular electrically charged rotating black holes and spinning
solitons (more precisely the Coleman lumps which are non-singular
non-dissipative objects holding themselves together by
self-interaction \cite{coleman}).


The key point is that for any gauge-invariant Lagrangian ${\cal
L}(F)$, stress-energy tensor of  electromagnetic field
 $$
\kappa T^{\mu}_{\nu}=-2{\cal L}_F
F_{\nu\alpha}F^{\mu\alpha}+\frac{1}{2}\delta^{\mu}_{\nu}{\cal
L};~\kappa=8\pi G; ~ \mu,\nu =0,1,2,3
                                                                            \eqno(3)
  $$
  where $F_{\mu\nu}=\partial_{\mu}A_{\nu}-\partial_{\nu}A_{\mu}$
  and ${\cal L}_F=d{\cal L}/dF$, in
  the spherically symmetric case
has the algebraic structure
 $$
T^t_t=T^r_r ~ ~ ~(p_r=-\rho) .
                                            \eqno(4)
  $$
Regular spherically symmetric solutions with stress-energy tensors
of any origin  specified by (4) and satisfying the weak energy
condition (non-negativity of density as measured by any local
observe), have obligatory de Sitter center with $p=-\rho$
\cite{me92,me2000,me2002} where $p=p_r=p_{\perp}$ and
$p_{\perp}=-\rho-r\rho^{\prime}/2$. In NED-GR regular solutions
interior de Sitter vacuum provides  a proper cut-off on
self-interaction divergent for a point charge
\cite{me2004,me2006}.

Nonlinear electrodynamics minimally coupled to gravity is
described by the action
 $$
S=\frac{1}{16\pi G}\int{d^4 x\sqrt{-g}[R-{\cal L}(F)]}; ~~ ~
F=F_{\mu\nu}F^{\mu\nu} ,
                                                          \eqno(5)
 $$
where $R$ is the scalar curvature. The Lagrangian ${\cal L}(F)$
should have the Maxwell limit, ${\cal L} \rightarrow F,~ {\cal
L}_F\rightarrow 1$ in the weak field regime. Variation  with
respect to $A^{\mu}$ and $g_{\mu\nu}$ yields
 the dynamical equations
 $$
  \nabla_{\mu}({\cal L}_F
F^{\mu\nu})=0;~~~ \nabla_{\mu}{^{\star}}F^{\mu\nu}=0  .
                                                       \eqno(6)
 $$
 where $^{\star}F^{\mu\nu}=\eta^{\mu\nu\alpha\beta}F_{\alpha\beta}/2;
~~ \eta^{0123}=-1/{\sqrt{-g}}$, and the Einstein equations
$G_{\mu\nu}=-\kappa T_{\mu\nu}$ with an electromagnetic source
$T_{\mu\nu}$ given by (3).

NED-GR equations do not admit regular spherically symmetric
solutions with the Maxwell center \cite{kirill}, but they  admit
regular solutions with the de Sitter center \cite{me2004}. The
question of correct description  of NED-GR regular electrically
charged structures in the frame of the Lagrange dynamics  is
clarified in \cite{us2015}.

Regular spherically symmetric solutions satisfying (4) are
described by the metric
 $$
 ds^2=g(r) dt^2 - \frac{dr^2}{g(r)} - r^2 d\Omega^2; ~g(r)=1-\frac{2{\cal M}(r)}{r}  ,
                                                                        \eqno(7)
 $$
 where
 ${\cal M}(r)=4\pi\int_0^r{~\tilde\rho(x)x^2dx}$ is calculated
 with the electromagnetic density $\tilde\rho(r)=T^t_t(r)$ from
(3). The metric (7) has the de Sitter asymptotic as $r\rightarrow
0$ and the Reissner-Nordstr\"om asymptotic as $r\rightarrow\infty$
\cite{me2004}.

The regular spherical solutions generated by (4) belong
to the Kerr-Schild class \cite{ks,behm,ssqv} and can be
transformed by the G\"urses-G\"ursey algorithm \cite{gurses} into
regular axially symmetric solutions which describe regular
rotating electrically charged objects, asymptotically Kerr-Newman
for a distant observer  \cite{burhild,me2006}.

In the Boyer-Lindquist coordinates the rotating metric reads
\cite{gurses}
  $$
   ds^2 =
\frac{2f(r) - \Sigma}{\Sigma} dt^2  + \frac{\Sigma}{\Delta} dr^2
+\Sigma d\theta^2
 - \frac{4af(r)\sin^2\theta}{\Sigma}dt d\phi
$$
$$
+ +\biggl(r^2 + a^2 +
\frac{2f(r)a^2\sin^2\theta}{\Sigma}\biggr)\sin^2\theta
d\phi^2;~\Sigma^2=r^2+a^2\cos^2\theta ,
                                                                                \eqno(8)
  $$
 where $\Delta = r^2 + a^2 - 2f(r)$.
 A function $f(r)=r{\cal M}(r)$ comes from a spherically
symmetric solution \cite{gurses}.   Regular spherical
 NED-GR solutions satisfy the weak energy condition, since
 $\kappa(p_{\perp}+\rho)=-F{\cal L}_F$, $F\leq 0$ and ${\cal
 L}_F\geq 0$ \cite{me2004}. In consequence,
${\cal M}(r)$ is non-negative function   growing from
$4\pi\tilde\rho(0)r^3/3$ as $r\rightarrow 0$ to $m-e^2/2r$ as
$r\rightarrow\infty$ \cite{me2004}. This guarantees the causal
safety on the whole manifold due to $f(r)\geq 0$ and $g_{\phi\phi}
> 0$ in (8).

The coordinate $r$ is defined as an affine parameter along either
of two principal null congruences, and
 the surfaces of constant $r$ are the oblate
confocal ellipsoids
 $$
r^4-r^2(x^2+y^2+z^2-a^2)-a^2 z^2=0 ,
                                                                            \eqno(9)
$$
 which degenerate, for $r = 0$, to the equatorial disk
 $$
 x^2 + y^2 \leq a^2, ~ ~~ z = 0 ,
                                                                           \eqno(10)
 $$
centered on the symmetry axis and bounded by the ring
$x^2+y^2=a^2$ ($r=0, \theta=\pi/2$)  \cite{chandra}.

Rotation transforms the de Sitter center to the de Sitter disk
(10). In the co-rotating  frame \cite{behm}
 the eigenvalues of the stress-energy tensor
are related to the function $f(r)$ as $\kappa\Sigma^2\rho = {2(f'r
- f); ~ ~\kappa\Sigma^2 p_{\perp} = 2(f'r - f) -
f^{\prime\prime}\Sigma}$ \cite{behm}. This gives
 $$
 \kappa\rho(r, \theta)=\frac{r^4}{\Sigma^2}{\tilde\rho}(r);
\kappa(p_{\perp}+\rho)=2\left(\frac{r^4}{\Sigma^2}-\frac{r^2}{\Sigma}\right)
{\tilde\rho}(r)-\frac{r^3}{2\Sigma}{\tilde\rho}^{\prime}(r) ,
                                                                               \eqno(13)
  $$
where $\tilde{\rho}(r)$ is a relevant spherically symmetric
density profile.  In the limit $r\rightarrow 0$, on the disk (10),
$r^2/\Sigma\rightarrow 1$ \cite{me2006}. For the spherical
solutions regularity requires
$r{\tilde\rho}^{\prime}(r)\rightarrow 0$ as $r\rightarrow 0$
\cite{me2004}. As a result we obtain on the disk the equation of
state
  $$
p_{\perp}+\rho=0  ~~~\rightarrow ~ p_{\perp}=p_r=-\rho ,
                                                                         \eqno(14)
  $$
which represents   the rotating de Sitter vacuum \cite{me2006}.

In terms of the  3-vectors of the electric induction ${\vec{D}}$
and magnetic induction $\vec{B}$, defined as
   $$
{E}_j=\{F_{j0}\},  {D}^j=\{{\cal L}_F F^{0j}\},
{B}^j=\{{^*}F^{j0}\},
 {H}_j=\{{\cal L}_F {^*}F_{0j}\},
                                                                            \eqno(15)
 $$
where $j,k =1,2,3$, the field equations (6) take the form of the
Maxwell equations. The inductions ${\vec{D}}$ and   $\vec{B}$ are
connected with the electric and magnetic field intensities by
   $$
D^{j}=\epsilon^{j}_{k}E^{k}; ~~ B^{j}=\mu^{j}_{k}H^{k} ,
                                                                                 \eqno(16)
 $$
where $\epsilon_{j}^{k}$ and $\mu_{j}^{k}$ are the tensors of the
electric and magnetic permeability \cite{me2006}
  $$
\epsilon_r^r=\frac{(r^2+a^2)}{\Delta}{\cal L}_F;
\epsilon_{\theta}^{\theta}={\cal L}_F;
\mu_r^r=\frac{(r^2+a^2)}{\Delta{\cal L}_F};
\mu_{\theta}^{\theta}=\frac{1}{{\cal L}_F}  .
                                                                                   \eqno(17)
  $$
The dynamical equations (6) are satisfied by
\cite{me2006,interior}
  $$
F_{10}=\frac{q}{\Sigma^2{\cal L}_F}(r^2-a^2\cos^2\theta);~
 F_{02}=\frac{q}{\Sigma^2{\cal L}_F}a^2r\sin 2\theta;
$$
$$
F_{31}=a\sin^2\theta F_{10}; ~ aF_{23}=(r^2+a^2)F_{02}
                                                                        \eqno(18)
 $$
 in the limit ${\cal L}_F\rightarrow\infty$, and
in the weak field limit ${\cal L}_F=1$ where they coincide with
the Kerr-Newman fields  \cite{carter,tiomno} and an integration
constant $q$ is identified as the electric charge.

 The relation connecting density and pressure with the electromagnetic
fields   reads \cite{me2006}
  $$
   \kappa (p_{\perp}+\rho)=2{\cal L}_F\left(
F_{10}^2+\frac{F_{20}^2}{a^2\sin^2\theta}\right) .
                                                                                      \eqno(19)
  $$
The field functions  (18)  give on the disk \cite{me2006,interior}
  $$
{\cal L}_F\Sigma^2=\frac{2q^2}{(p_{\perp}+\rho)};
~~F=-\frac{\kappa^2(p_{\perp}+\rho)^2\Sigma^2}{2q^2} .
                                                                                 \eqno(20)
  $$
Equation of state on the disk (14)  requires ${\cal
L}_F\Sigma^2\rightarrow\infty$ and hence ${\cal
L}_F\rightarrow\infty$ faster than $\Sigma^{-2}$. The electric
permeability in (17) goes to infinity, the magnetic permeability
vanishes, so that the de Sitter vacuum disk has both perfect
conductor and ideal diamagnetic properties. The magnetic induction
$\mathbf{B}$ vanishes on the disk \cite{me2006}.

In electrodynamics of continued media the transition to a
superconducting state corresponds to the limits
${\mathbf{B}}\rightarrow 0$ and $\mu\rightarrow 0$ in a surface
current  ${\mathbf{j_s}}=
\frac{(1-\mu)}{4\pi\mu}[{\mathbf{n}}{\mathbf{B}}]$, where
${\mathbf{n}}$ is the normal to the surface. The right-hand side
then becomes indeterminate, and there is no condition which would
restrict the possible values of the current \cite{landau2}. On the
de Sitter disk we can apply definition of a surface current for a
charged surface layer, $4\pi
j_k=[e_{(k)}^{\alpha}F_{\alpha\beta}n^{\beta}]$ \cite{werner},
where $[..]$ denotes a jump across the layer; $e_{(k)}^{\alpha}$
are the tangential base vectors associated with the intrinsic
coordinates on the disk  $t,\phi$, $0\leq\xi\leq\pi/2$ labeled by
k=1,2,3; $n_{\alpha}=(1+q^2/a^2)^{-1/2}\cos\xi~\delta^1_{\alpha}$
is the unit normal directed upwards \cite{werner}. With using
asymptotic solutions (18) and magnetic permeability from (17),
$\mu^r_r=\mu^{\theta}_{\theta}=\mu=1/{\cal L}_F\rightarrow 0$, we
obtain the surface current
   $$
   j_{\phi}=-\frac{q}{2\pi a}
   ~\sqrt{1+q^2/a^2}~\sin^2\xi~\frac{\mu}{\cos^3\xi} .
                                                                  \eqno(21)
 $$
At approaching the ring $r=0,~\xi=\pi/2$, both terms in the second
fraction go to zero independently. As a result the surface
currents on the ring can be any and amount to a non-zero total
value.

\vskip0.1in

In terms of a density $\tilde{\rho}(r)$ and pressure of a related
spherical solution,
$\tilde{p_{\perp}}=-\tilde\rho-r\tilde\rho^{\prime}/2$
\cite{me2004},
$$
  \kappa(p_{\perp}+\rho)=\frac{r|\tilde\rho^{\prime}|}{2\Sigma^2}~{\cal
  E}(r,z)=(r^4-z^2P(r)), P(r)
  =\frac{2a^2}{r|\tilde\rho^{\prime}|}(\tilde\rho
  -\tilde{p_{\perp}}) .
                                                    \eqno(22)
  $$
This implies a possibility of generic violation of the weak energy
condition (WEC) which was reported for several regular rotating
solutions \cite{behm,neves,bambi,stuchlik}. WEC can be violated
beyond a de Sitter vacuum surface ${\cal E}(r, z)=0$ defined by
$p_{\perp}+\rho=0$ where the right-hand side in (22) can change
sign provided that the dominant energy condition ($\tilde\rho \geq
\tilde{p_k}$) is valid for related spherical solutions.

Each point of the ${\cal E}$-surface belongs to some of confocal
ellipsoids (9) covering the whole space as the coordinate surfaces
$r=const$. In the Cartesian coordinates, $
 x^2 + y^2 =
(r^2 + a^2)\sin^2\theta;~z=r\cos\theta$, the  squared width of the
${\cal E}$-surface,  $W^2_{\cal E}=(x^2+y^2)_{\cal
E}=(a^2+r^2)(1-z^2/r^2)_{\cal E}=(a^2+r^2)(1-r^2/P(r))$. For
regular solutions $r\tilde\rho^{\prime}\rightarrow 0$,~
$p_{\perp}\rightarrow -\rho$ as $r\rightarrow 0$ \cite{me2004},
and $P(r)\rightarrow A^2r^{-(n+1)}$ with the integer $n\geq 0$ as
$r\rightarrow 0$. Then $W^2_{\cal E} =(a^2+r^2)(1-r^{(n+3)}/A^2)$,
and $W^2\rightarrow a^2$ when $r\rightarrow 0$. As a function of
$z$, $W^2_{\cal E}=(x^2+y^2)_{\cal E} =
(a^2+|z|\sqrt{P(r)})(1-|z|/\sqrt{P(r)})$, and ${\cal E}$-surface
is entirely confined within the $r_*$-ellipsoid whose minor axis
coincides with $|z|_{max}$ for the ${\cal E}$-surface
\cite{interior}.  The derivative of $W_{\cal E}(z)$ near
$z\rightarrow 0$ behaves as $z^{-(n+1)/(n+5)}$ and goes to
$\pm\infty$ as $z\rightarrow 0$, so that the function $W_{\cal
E}(z)$ has the cusp at approaching the disk and two symmetric
maxima between $z=\pm r_*$ and $z=0$ \cite{interior}.

In Fig.1 \cite{interior} ${\cal E}$- surface is plotted  for the
regularized Coulomb profile \cite{me2004}
 $$
  \tilde\rho=\frac{q^2}{(r^2+r_q^2)^2};~~r_q=\frac{\pi q^2}{8 m} .
                                                                           \eqno(23)
 $$
Its width in the equatorial plane $W_{\cal E}=a$ and the height
$H_{\cal E} =|z|_{max}=\sqrt{ar_q}$. Its precise form depends on
the relation of two parameters, $\alpha=a/m$ and $\beta=q/m$. For
black holes the parameter $\beta$ changes within the range $0 <
\beta < 0.99$ \cite{cardoso,hamilton}. For $\alpha <
(\pi/8)\beta^2$ the ${\cal E}$-surface is prolate. It can be the
case for a slowly rotating moderately charged black hole. For
$\alpha > (\pi/8)\beta^2$, the ${\cal E}$-surface, shown in Fig.1,
is oblate. It can be the case of slightly charged rotating black
holes and extreme black holes. It is also the case of
electromagnetic soliton (e-lump).
\begin{figure}[htp]
\centering \epsfig{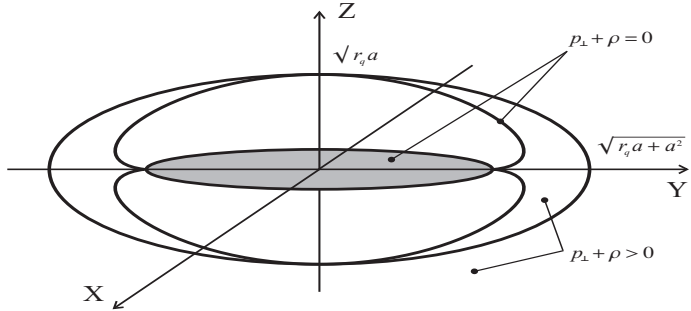}
\caption{${\cal E}$-surface for the case $\alpha >
(\pi/8)\beta^2$.}
 \label{fig}
\end{figure}

${\cal E}$-surface is defined by $p_{\perp}+\rho=0$, it follows
${\cal L}_F\rightarrow\infty$ by virtue of the first expression in
(20).  The magnetic permeability vanishes and electric
permeability goes to infinity, so that ${\cal E}$-surface displays
the properties of a perfect conductor and ideal diamagnetic.

The de Sitter vacuum ${\cal E}$-surface contains de Sitter disk as
a bridge. Magnetic induction vanishes throughout  the whole
surface by virtue of (15) with taking into account (17) and (18).
Within ${\cal E}$-surface, in cavities between its upper and down
boundaries and the bridge, a negative value of $(p_{\perp}+\rho)$
in (19)~
 would mean  negative values for electric and magnetic
 permeabilities in (17) inadmissible in electrodynamics of continued media
 \cite{landau2}.

The alternative compatible with regularity
 is zero value of $(p_{\perp}+\rho)$ also inside ${\cal E}$-surface.
 This could be the case for the shell-like models
 (\cite{lopez} and references therein) and baglike models (\cite{bur2015} and references therein)
 with the flat vacuum
 interior, zero interior fields and in consequence zero density and
 pressures.

 The other possibility, favored by the underlying idea of nonlinearity replacing
 a singularity and suggested by vanishing of magnetic induction on the surrounding
${\cal E}$-surface, is  extension of ${\cal L}_F\rightarrow\infty$
to its interiors. This results in $p_{\perp}=-\rho$ and we have de
Sitter vacuum core
 with the properties of a perfect conductor and ideal diamagnetic
 and vanishing magnetic induction.

\vskip0.1in

The ${\cal E}$-surface exists in the case when a related spherical
solution satisfies the dominant energy condition. In the opposite
case $p_{\perp}+\rho \geq 0$ throughout the whole manifold. The de
Sitter vacuum disk $p_{\perp}+\rho=0$, confined by the de Sitter
ring replacing the Kerr singularity, exists for any regular
rotating object.

 In any case superconducting currents flowing
 on the de Sitter vacuum ring can be considered as a source
 of the Kerr-Newman electromagnetic fields. This kind of a source is non-dissipative
 so that life time of a spinning electrically charged object, in particular
 of the electron, can be practically unlimited.

For the electron $q=-e$, $ma=\hbar/2$ \cite{carter},
$a=\lambda_e/2$, where $\lambda_e =\hbar/(m_ec)$ is the Compton
wavelength. In the observer region $r\gg\lambda_e$
   $$
   E_r=-\frac{e}{r^2}\left(1-\frac{\hbar^2}{m_e^2c^2}\frac{3\cos^2\theta}{4r^2}\right);~
   E_{\theta}=\frac{e\hbar^2}{m_e^2c^2}\frac{\sin 2\theta}{4r^3};
                                                                                         \eqno(24a)
   $$
   $$
    B^r=-\frac{e\hbar}{m_ec}\frac{\cos\theta}{r^3}=2\mu_e\frac{\cos\theta}{r^3};
   ~ B_{\theta}=-\mu_e\frac{\sin\theta}{r^4} .
                                                                              \eqno(24b)
 $$
The Planck constant appears here due to discovered by Carter
ability of the Kerr-Newman solution to present the electron as
seen by a distant observer. In terms of the Coleman lump eq. (24)
describes the following situation: The leading term in $E_r$ gives
the Coulomb law as the classical limit $\hbar=0$, the higher terms
represent the quantum corrections.


\end{document}